\documentclass[pra,showkeys,twocolumn]{revtex4}
\usepackage{amsmath}
\usepackage{amscd}
\usepackage{graphicx}
\usepackage{subfigure}
\usepackage[toc,page,title,titletoc,header]{appendix}
\usepackage{amsfonts}
\usepackage{multirow}
\usepackage{bm}
\usepackage{bbm}

\begin{document}

\title[Short Title]{Invariant-based inverse engineering for fluctuation transfer between membranes in an optomechanical cavity system}

\author{Ye-Hong Chen$^{1,2}$}
\author{Zhi-Cheng Shi$^{1,2,}$\footnote{E-mail: szc2014@yeah.net}}
\author{Jie Song$^{3}$}
\author{Yan Xia$^{1,2,}$\footnote{E-mail: xia-208@163.com}}

\affiliation{$^{1}$Department of Physics, Fuzhou University, Fuzhou 350002, China\\
             $^{2}$Fujian Key Laboratory of Quantum Information and Quantum Optics (Fuzhou University), Fuzhou 350116, China\\
             $^{3}$Department of Physics, Harbin Institute of Technology, Harbin 150001, China}


\begin{abstract}
  In this paper, by invariant-based inverse engineering, we design classical driving fields to transfer quantum fluctuations between two suspended membranes in an optomechanical cavity system.
  The transfer can be quickly attained through a non-adiabatic evolution path determined by a so-called
  dynamical invariant. Such an evolution path allows one to
  optimize the occupancies of the unstable ``intermediate'' states thus the influence of cavity decays can be suppressed.
  Numerical simulation demonstrates that a perfect fluctuation transfer between two membranes
  can be rapidly achieved in one step, and the transfer is robust to both the amplitude noises and cavity decays.
\end{abstract}

\pacs {03.67. Pp, 03.67. Mn, 03.67. HK} \keywords{Invariant-based inverse engineering; Fluctuation transfer; Optomechanical system}

\maketitle
\section{Introduction}
The field of quantum optomechanics has been intensively investigated
due to its fundamental aspects in quantum mechanics of macroscopic
bodies and possible applications in quantum metrology and hybrid
quantum systems \cite{Sci3211172,Rmp861391}. It explores the
interaction between light and mechanical motion by composing an
optical (or microwave) cavity and a mechanical resonator. Over the
past decades, tremendous progress has been made in the field of
quantum optomechanics. On the one hand, nontrivial quantum
phenomena have been realized in optomechanical
systems, such as sideband \cite{Nat475359,Prl99093902}
and near-ground-state cooling \cite{Nat464697,Nat47889,Prl99093901,Prl110153606}, strong
coupling effects \cite{Nat460724,Nat471204,Prl114093602}, squeezing of a
mechanical oscillator \cite{Prl107213603,Pra79063819,Pra89023849},
and so on \cite{Pra82021806,Nat48263,Prl109013603}.
On the other hand, complementary setups have been
realized, for example, cavities with an oscillating end mirror
\cite{Np8168,Prl101133903}, evanescent-wave resonators \cite{Np5909},
and membrane-in-the-middle (MIM) cavities \cite{Nat45272,Prl103207204,Njp16033023}.
In recent years, because optomechanical systems have the ability to
transfer a quantum state between photons with vastly
differing wavelengths (quantum information and quantum fluctuations of
an optical field can be reversibly mapped to a mechanical
state), the direct utility of such systems in quantum information
processing becomes attractive \cite{Nat45272,Njp13013017,Pra84043845,
Pra82053806,Prl105220501,Nat459546,Pra88033802,Prl107133601,Prl110253601,Pra88022325,Pra92013852,Pra88033614,Nc711736}.
For example, Tian and Wang have studied a quantum state conversion between
cavity modes of distinctly different wavelengths by applying
classical driving fields to swap the cavity and the mechanical states \cite{Pra82053806}.
Fiore \emph{et al.} have experimentally reported the demonstration of storing optical information as a mechanical excitation
in a silica optomechanical resonator \cite{Prl107133601}.


Noting that, in the process of exploiting the utility of
optomechanical systems for quantum information processing,
growing interest has been shown toward the adiabatic control of such systems
because a dark-state evolution is usually robust against noise and dissipation
\cite{Prl108153603,Pra96023837,Sci3381609,Prl108153604}.
For example, Wang and Clerk performed the high-fidelity transfer of a quantum
state between two electromagnetic cavities by adiabatic control \cite{Prl108153603}.
Soon after that, Dong \emph{et al.} experimentally realized an
adiabatic transfer of optical fields between two optical modes of a
silica resonator \cite{Sci3381609}. Recently, Garg \emph{et al.} proposed a scheme to
transfer quantum fluctuations between mechanical oscillators based on adiabatic control \cite{Pra96023837}.
These researches further open up the possibility of using optomechanical coupling in various
applications without cooling the mechanical oscillator to its ground
state. However, applying adiabatic control usually requires a
long-time interaction to guarantee the adiabaticity of the system.
To speed up the adiabatic schemes without losing the advantages of adiabatic control,
various versions of approaches called
shortcuts to adiabaticity (STA)
\cite{Prl105123003,ETSISMGMMACDGOARXCJGMAmop13,Prl111100502,Prl109100403}
have been proposed in recent years, including, e.g., transitionless driving algorithm
\cite{Jpca1079937,Jpa42365303,Oc13948,Epjst224189,Pra82033430,Epl9323001},
inverse engineering based on invariants
\cite{Jpb42241001,Prl104063002,Pra86033405,Pra89053408,Pra89043408,Pra89033856,Njp13113017,Pra83013415,Njp14013031,Pra83043804}, and so on
\cite{Jpb43085509,Pra84031606Epl9660005,arXiv160105551,Njp16015025,Prl116230503,Natc712479,Pra93052324,Pra93052109}.
The basic idea of the STA approach is to design a suitable Hamiltonian
to drive a quantum system to evolve along a non-adiabatic route to
reproduce the same final state of an adiabatic process. Thus, the
relevant adiabatic requirement can be removed and the evolution
could be driven fast. Very recently, the STA approach has been applied to
 optomechanical systems for fast quantum state
conversions \cite{Lpl14095202,arXiv170307933}.

In this paper, we proceed further to exploit the use of STA in
optomechanical systems. Precisely speaking, we show how to use
inverse engineering based on Lewis-Riesenfeld invariants to coherently and
deterministically transfer the quantum fluctuations from one
mechanical oscillator to the other.
The invariant-based inverse engineering is an efficient non-adiabatic approach (usually based on the Lewis-Riesenfeld theory) to analytically
obtain an exact dynamical evolution for an arbitrary quantum system. The Lewis-Riesenfeld theory \cite{Jmp101458}
tells that the quantum dynamics of a quantum system can be dictated by a dynamical invariant $I(t)$ obeying the von Neumann equation
$\frac{d}{dt}I(t)\equiv {\partial_t}I(t)-{i}[H(t),I(t)]=0$ (setting $\hbar=1$).
An arbitrary solution of a time-dependent Schr\"{o}dinger equation
$i{\partial_t}|{\Psi}(t)\rangle=H(t)|\Psi(t)\rangle$ can be
expressed as $|\Psi(t)\rangle=\sum_{n}C_{n}e^{i\eta_{n}(t)}|\phi_{n}(t)\rangle$,
where $C_{n}$ is a time-independent amplitude, $|\phi_{n}(t)\rangle$
is the $n$th eigenvector of the invariant $I(t)$, and $\eta_{n}$ is
the corresponding Lewis-Riesenfeld phase,
\begin{align}\label{eq0-1}
  \eta_{n}=\int_{t_{i}}^{t}\langle\phi_{n}(t')|{\partial_t'}-H(t')|\phi_{n}(t')\rangle dt'.
\end{align}
Note that one should impose $[I(t_{i}),H(t_{i})]\simeq0$ and
$[I(t_{f}),H{(t_{f})}]\simeq0$ to ensure that $I(t)$ and $H(t)$ share the
same eigenvectors at the initial and final times. In this case,
an exact dynamical evolution can be obtained by only designing the parameters for a Hamiltonian $H(t)$,
thus the scheme would be easily realized in practice.
Over the past decades, the invariant-based inverse engineering has been widely devoted to achieve
a short-time adiabatic-like evolution and applied in trap
expansions \cite{Jpb42241001,Prl104063002}, rotations
\cite{Njp13113017}, atom transport \cite{Pra83013415,Njp14013031},
and mechanical oscillators \cite{Pra83043804}.

By applying the inverse engineering based on Lewis-Riesenfeld invariants,
we show how quantum fluctuations can be transferred from one mechanical
oscillator to the other in a short time with a high fidelity.
Specifically, we consider a weak optomechanical coupling to apply the rotating-wave
approximation and quantum Zeno dynamics \cite{Prl89080401Jpa41493001} to obtain an effective Hamiltonian.
Then, in absence of decays of the cavity modes and the membranes, we design parameters by invariant-based inverse engineering to complete the fluctuation transfer.
This is different from the
previous reported works \cite{Lpl14095202,arXiv170307933} wherein such
a transfer occurs between two cavity modes. The exchange of the
energy fluctuations between two mechanical systems may pose as
a possible quantum communication protocol \cite{Pra96023837,Pra82053806}.
In this scheme, the occupancies of the unstable ``intermediate'' states can be optimized by suitably choosing the parameters that suppress the cavity decays.
The sensitivity with respect to amplitude-noise errors in the driving fields is also analyzed by numerical simulation.
The result shows the system is robust against amplitude noises with suitable parameters.

The paper is organized as follows. In Sec. II, we present the model and the corresponding effective Hamiltonian.
In Sec. III, we show how to design the Hamiltonian by invariant-based inverse engineering to transfer quantum fluctuations.
In Sec. IV, we analyze the robustness of the scheme against noises and decays by numerical simulation.
The conclusion is given in Sec. V.

\section{Model}

\begin{figure}[t]
 \scalebox{0.25}{\includegraphics {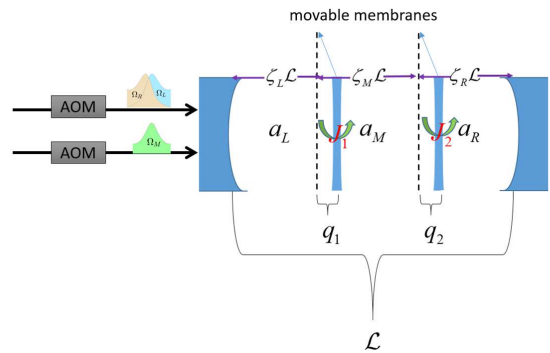}}
 \caption{
         Schematic diagram. Two membranes are suitably placed and the entire cavity is thus divided into three subcavities (marked as L, M, and R, respectively).
         The length of the cavity is $\mathcal{L}$ and the length of the $j$th subcavity is $\zeta_{j}\mathcal{L}$, where $\zeta_{j}=\frac{\omega_{j}}{\Xi}$ and $\Xi=\sum_{j}\omega_{j}$ ($j=L,R,M$).
         Caused by the optical radiation pressure, the membranes 1 and 2 will move with displacements $q_{1}$ and $q_{2}$, respectively.
         The acousto-optic modulators (AOMs) are used to modulate the shapes of the classical driving fields.
         Assuming the mass of the $k$th membrane is $\mu_{k}$, we find the optomechanical couplings $g_{1}$ and $g_{2}$ in Eq. (\ref{eq1-1}) are $g_{1}=\frac{\Xi}{\mathcal{L}}\sqrt{\frac{1}{2\mu_{1}\omega_{m,1}}}$ and $g_{2}=\frac{\Xi}{\mathcal{L}}\sqrt{\frac{1}{2\mu_{2}\omega_{m,2}}}$ (see Appendix A for details), respectively.
         }
 \label{fig1}
\end{figure}
As shown in Fig. \ref{fig1}, we consider an optomechanical cavity system with two suspended membranes.
Assuming that the membranes are fully reflecting at their surfaces, the system can be divided into three independent subcavities.
For convenience, we mark the three subcavities as: left (L), middle (M), and right (R), respectively.
In this case, the $k$th ($k=1,2$) membrane is partially transmitting, one
can have a tunneling between the two subcavities on either side
of the membrane with a rate $J_{k}$. Then, sending three monochromatic waves $\Omega_{L}$, $\Omega_{R}$, and $\Omega_{M}$ whose polarizations are orthogonal to each other
along the cavity axis to drive the modes $a_{L}$, $a_{R}$, and $a_{M}$, respectively. We choose the same angular frequency $\omega_{l}$ for classical driving fields $\Omega_{L}$ and $\Omega_{R}$,
while choose a different frequency $\omega_{l}'$ for the classical driving field $\Omega_{M}$ in order to avoid any crosstalk \cite{Sci3381609}.
Beware that each of the driving fields we used in this paper is a monochromatic wave with a specific frequency, which is different from a ``real'' pulse with several frequencies \cite{Nc815886}.
The Hamiltonian \cite{Pra512537,Pra87013839,Pra96023837,Prl109063601,Prl103100402,Njp16033023,Pra92013822,Pra77033819,Prl109013603} for the system reads ($\hbar=1$)
\begin{align}\label{eq1-1}
  H=&H_{0}+H_{I}+H_{J}+H_{D}, \cr
  H_{0}=&\sum_{j=L,M,R}{\omega_{c,j}a_{j}^{\dag}a_{j}}+\sum_{k=1,2}\omega_{m,k}b_{k}^{\dag}b_{k}, \cr
  H_{I}=&-g_{1}(a_{L}^{\dag}a_{L}-a_{M}^{\dag}a_{M})(b_{1}+b_{1}^{\dag})\cr
        &+g_{2}(a_{R}^{\dag}a_{R}-a_{M}^{\dag}a_{M})(b_{2}+b_{2}^{\dag}), \cr
  H_{J}=&-(J_{1}a_{L}^{\dag}a_{M}+J_{2}a_{M}^{\dag}a_{R}+H.c.), \cr
  H_{D}=&\Omega_{L}a_{L}e^{i\omega_{l}t}+\Omega_{R}a_{R}e^{i\omega_{l}t}\cr
        &+\Omega_{M}a_{M}e^{i\omega_{l}'t}+H.c.,
\end{align}
where $a_{j}$ ($j=L,M,R$) is the annihilation operator for the $j$th cavity mode with corresponding
frequency $\omega_{c,j}$, $b_{k}$ ($k=1,2$) is the annihilation operator of the $k$th mechanical mode with
respective frequency $\omega_{m,k}$, $g_{1}$ ($g_{2}$)
determines the optomechanical coupling for the first (second)
membrane with left (right) and middle modes, while $J_{1}$ ($J_{2}$)
represents the transmission coefficient between the left (right)
and middle cavity modes through first (second) membrane.
Beware that $J_{k}=0.5\omega_{m,k}$ is necessary to ensure that the coupling is linear in the displacement
quadrature $X\equiv(b+b^{\dag})$ in a MIM setup.
In the rotating frame of laser frequencies, the Hamiltonian $H$ takes the following form,
\begin{align}\label{eq1-2}
  H=&\sum_{j=L,M,R}{[\Delta_{j}a_{j}^{\dag}a_{j}+\Omega_{j}(a_{j}+H.c.)]}+\sum_{k=1,2}{\omega_{m,k}b_{k}^{\dag}b_{k}} \cr
    &-g_{1}(a_{L}^{\dag}a_{L}-a_{M}^{\dag}a_{M})(b_{1}+b_{1}^{\dag}) \cr
    &+g_{2}(a_{R}^{\dag}a_{R}-a_{M}^{\dag}a_{M})(b_{2}+b_{2}^{\dag}) \cr
    &-(J_{1}a_{L}^{\dag}a_{M}+J_{2}a_{M}^{\dag}a_{R}+H.c.),
\end{align}
where $\Delta_{j}=\omega_{c,j}-\omega_{l}$ and $\Delta_{M}=\omega_{c,M}-\omega_{l}'$.
By using the standard linearization procedure \cite{Rmp861391}, all the bosonic operators
can be expanded as a sum of
the average values and the zero-mean fluctuation as follows:
$a_{j}\rightarrow \alpha_{j}+\delta a_{j}$, and $b_{k}\rightarrow \beta_{k}+\delta b_{k}$,
where $\alpha_{j}$ and $\beta_{k}$ are generally complex and denote the steady-state values of the respective
annihilation operators. Then, an effective Hamiltonian for the operator fluctuations (derivation is given in Appendix B) is given as
\begin{align}\label{eq1-3}
  H=&-g_{1}(\alpha_{L}^{*}\delta b_{1}^{\dag}\delta a_{L}-\alpha_{M}^{*}\delta b_{1}^{\dag}\delta a_{M}) \cr
    &+g_{2}(\alpha_{R}^{*}\delta b_{2}^{\dag}\delta a_{R}-\alpha_{M}^{*}\delta b_{2}^{\dag}\delta a_{M}) \cr
    &-(J_{1}\delta a_{L}^{\dag}\delta a_{M}+J_{2}\delta a_{R}^{\dag}\delta a_{M})+H.c.,
\end{align}
where
\begin{align}\label{eq1-4}
  \alpha_{L}=&\frac{\Omega_{L}-J_{1}\alpha_{M}}{-\Delta_{0}+i\gamma_{L}/2}, \cr
  \alpha_{R}=&\frac{\Omega_{R}-J_{2}\alpha_{M}}{-\Delta_{0}+i\gamma_{R}/2}, \cr
  \alpha_{M}=&\frac{\Omega_{M}-J_{1}\alpha_{L}-J_{2}\alpha_{R}}{-\Delta_{0}+i\gamma_{M}/2}.
\end{align}
Here $\Delta_{0}$ is given according to Eq. (\ref{eqa-9}) and $\gamma_{j}$ is the decay rate of the $j$th cavity mode.
For the sake of simplification, we choose $\Omega_{M}=J_{1}\alpha_{L}+J_{2}\alpha_{R}$ so that
\begin{align}\label{eq1-5}
  \alpha_{M}&=0,\cr \alpha_{L}&=\frac{\Omega_{L}}{-\Delta_{0}+i\gamma_{L}/2},\cr \alpha_{R}&=\frac{\Omega_{R}}{-\Delta_{0}+i\gamma_{R}/2}.
\end{align}

Using the Hamiltonian in Eq. (\ref{eq1-3}) and according to the input-output formalism \cite{Quantumoptics}, we obtain Langevin equations for the
operator fluctuations as \cite{Pra96023837}
\begin{align}\label{eq1-6}
  \delta \dot{a}_{L}=&i J_{1}\delta a_{M}+i g_{1}\alpha_{L} \delta b_{1}-\frac{\gamma_{L}}{2}\delta a_{L}+\sqrt{\gamma_{L}} \delta a_{L}^{\text{in}}, \cr
  \delta \dot{a}_{R}=&i J_{2}\delta a_{M}-i g_{2}\alpha_{R} \delta b_{2}-\frac{\gamma_{R}}{2}\delta a_{R}+\sqrt{\gamma_{R}} \delta a_{R}^{\text{in}}, \cr
  \delta \dot{a}_{M}=&i J_{1}\delta a_{L}+i J_{2}\delta a_{R}-\frac{\gamma_{M}}{2}\delta a_{M}+\sqrt{\gamma_{M}} a_{M}^{\text{in}}, \cr
  \delta \dot{b}_{1}=&ig_{1}\alpha_{L}\delta a_{L}-\frac{\gamma_{m,1}}{2}\delta b_{1}+\sqrt{\gamma_{m,1}}\delta b_{1}^{in}, \cr
  \delta \dot{b}_{2}=&-ig_{2}\alpha_{R}\delta a_{R}-\frac{\gamma_{m,2}}{2}\delta b_{2}+\sqrt{\gamma_{m,2}}\delta b_{2}^{in}, \cr
\end{align}
where the overdot stands for a time derivative and $\delta a_{j}^{\text{in}}$ and $\delta b_{k}^{\text{in}}$ denote the fluctuations of the noise operators.
We can rewrite the Langevin equations in Eqs. (\ref{eq1-6}) in matrix form as $i{|\dot\Psi}(t)\rangle=M(t)|\Psi(t)\rangle$, where
\begin{align}\label{eq1-7}
  |\Psi(t)\rangle=\left[
        \begin{array}{c}
          \delta{a}_{L} \\
          \delta{a}_{R} \\
          \delta{a}_{M} \\
          \delta{b}_{1} \\
          \delta{b}_{2}
        \end{array}
  \right],
\end{align}
\begin{small}
\begin{align}
  M(t)=\left[
        \begin{array}{ccccc}
          -i\gamma_{L}/2 & -J_{1} & 0 & -g_{1}\alpha_{L} & 0 \\
          -J_{1} & -i\gamma_{M}/2 & -J_{2} & 0 & 0 \\
          0 & -J_{2} & -i\gamma_{R}/2 & 0 & g_{2}\alpha_{R} \\
          -g_{1}\alpha_{L} & 0 & 0 & -i\gamma_{m,1}/2 & 0 \\
          0 & 0 & g_{2}\alpha_{R} & 0 & -i\gamma_{m,2}/2
        \end{array}
  \right].
\end{align}
\end{small}
Assuming that the decay rates of the subcavities and membranes are much smaller than their
fundamental frequencies such that there is no significant decay of the photons during the transfer process,
the non-Hermitian terms in Eqs. (\ref{eq1-6}) can be neglected.

Referring to the formula of quantum Zeno dynamics \cite{Prl89080401Jpa41493001}, we write the matrix $M(t)$ in the absence of the decay terms ($\gamma_{j}=\gamma_{m,k}=0$) as
$M(t)=\Omega(M_{p}+KM_{q})$, where $\Omega=\sqrt{(g_{1}\alpha_{L})^2+(g_{2}\alpha_{R})^2}$, $K=\sqrt{2}J/\Omega$, and
\begin{align}\label{eq1-8}
  M_{p}=\frac{1}{\Omega}\left[
        \begin{array}{ccccc}
          0 & 0 & 0 & -g_{1}\alpha_{L} & 0 \\
          0 & 0 & 0 & 0 & 0 \\
          0 & 0 & 0 & 0 & g_{2}\alpha_{R} \\
          -g_{1}\alpha_{L} & 0 & 0 & 0 & 0 \\
          0 & 0 & g_{2}\alpha_{R} & 0 & 0
        \end{array}
  \right],
\end{align}
\begin{align}
  M_{q}=\frac{1}{\sqrt{2}}\left[
        \begin{array}{ccccc}
          0 & -1 & 0 & 0 & 0 \\
          -1 & 0 & -1 & 0 & 0 \\
          0 & -1 & 0 & 0 & 0 \\
          0 & 0 & 0 & 0 & 0 \\
          0 & 0 & 0 & 0 & 0
        \end{array}
        \right].
\end{align}
Here we have chosen $J_{1}=J_{2}=J$.
When the strong coupling limit $K\rightarrow\infty$ ($\sqrt{2}J\gg\Omega$) is satisfied,
we obtain the effective interaction matrix $M_{eff}=(\sum_{l}P_{l}M_{p}P_{l}+K\epsilon_{l}P_{l})$,
where $P_{l}$ is the $l$th eigenprojection and $\epsilon_{l}$ is the corresponding
eigenvalue of $M_{q}$: $M_{q}=\sum_{l}\epsilon_{l} P_{l}$.
Thus, in the Zeno dark subspace ($\epsilon_{l}=0$) spanned by
\begin{align}\label{eq1-9}
  |\psi_{1}\rangle=\left[
             \begin{array}{c}
               0 \\
               0 \\
               0 \\
               1 \\
               0
             \end{array}
       \right],
  |\psi_{2}\rangle=\left[
             \begin{array}{c}
               0 \\
               0 \\
               0 \\
               0 \\
               1
             \end{array}
       \right],
  |\psi_{3}\rangle=\frac{1}{\sqrt{2}}\left[
             \begin{array}{c}
               1 \\
               0 \\
               -1 \\
               0 \\
               0
             \end{array}
       \right],
\end{align}
the effective interaction matrix $M_{eff}(t)$ reads
\begin{align}\label{eq1-10}
  M_{eff}(t)=&-\frac{1}{\sqrt{2}}(g_{1}\alpha_{L}|\psi_{1}\rangle\langle\psi_{3}|+g_{1}\alpha_{L}|\psi_{3}\rangle\langle\psi_{1}| \cr
             &+g_{2}\alpha_{R}|\psi_{2}\rangle\langle\psi_{3}|+g_{2}\alpha_{R}|\psi_{3}\rangle\langle\psi_{2}|).
\end{align}

\section{invariant-based inverse engineering for fast energy fluctuation transfer}

We show in this section how to transfer fluctuation excitation from mode $b_{1}$ to mode
$b_{2}$ in a fast and robust way.
For the time-dependent matrix $M_{eff}(t)$ in Eq. (\ref{eq1-10}), the
dynamical invariant $I(t)$ satisfying the von Neumann-like equation ${\partial_{t}}I(t)-{i}[M_{eff}(t),I(t)]=0$
is given by \cite{Pra533691Jpa291773}
\begin{align}\label{eq1-11}
  I(t)=&\cos{\varphi}\sin{\theta}|\psi_{1}\rangle\langle\psi_{3}|+\cos{\varphi}\cos{\theta}|\psi_{2}\rangle\langle\psi_{3}|\cr
       &-i\sin{\varphi}|\psi_{1}\rangle\langle\psi_{2}|+\cos{\varphi}\sin{\theta}|\psi_{3}\rangle\langle\psi_{1}|\cr
       &+\cos{\varphi}\cos{\theta}|\psi_{3}\rangle\langle\psi_{2}|+i\sin{\varphi}|\psi_{2}\rangle\langle\psi_{1}|.
\end{align}
The parameters satisfy
\begin{align}\label{eq1-12}
  g_{1}\alpha_{L}=&-\sqrt{2}(\dot{\theta}\cot{\varphi}\sin{\theta}+\dot{\varphi}\cos{\theta}) \cr
                 =&-\Omega\sin{\varrho},\cr
  g_{2}\alpha_{R}=&-\sqrt{2}(\dot{\theta}\cot{\varphi}\cos{\theta}-\dot{\varphi}\sin{\theta}) \cr
                 =&-\Omega\cos{\varrho},
\end{align}
where
\begin{align}
  \Omega=&\sqrt{2[(\dot{\theta}\cot{\varphi})^{2}+\dot{\varphi}^{2}]},\cr
  \varrho=&\theta+\arctan(\frac{\dot{\varphi}}{\dot{\theta}\cot{\varphi}}).
\end{align}
The driving fields are thus designed according to Eqs. (\ref{eq1-5}) and (\ref{eq1-12}),
\begin{align}\label{eq1-12b}
  \Omega_{L}=&\frac{\sqrt{2}\Delta_{0}}{g_{1}}(\dot{\theta}\cot{\varphi}\sin{\theta}+\dot{\varphi}\cos{\theta}),\cr
  \Omega_{R}=&\frac{\sqrt{2}\Delta_{0}}{g_{2}}(\dot{\theta}\cot{\varphi}\cos{\theta}-\dot{\varphi}\sin{\theta}), \cr
  \Omega_{M}=&-\frac{J}{\Delta_{0}}(\Omega_{L}+\Omega_{R}).
\end{align}

\begin{figure}
 \scalebox{0.28}{\includegraphics {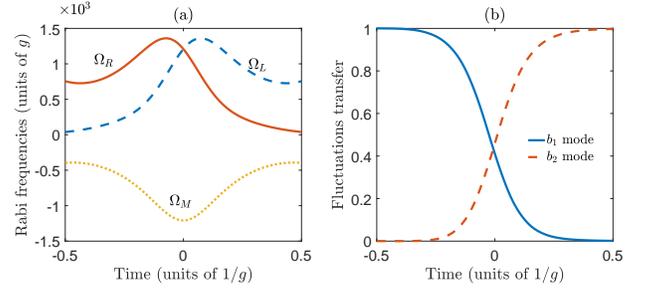}}
 \caption{
         (a) The classical driving fields given according to Eq. (\ref{eq1-12b}).
         We have chosen $g_{1}=g_{2}=g=2\pi\times10\text{KHz}$. Parameters are $T=1/g$, $\tau=0.1T$, $\tau_{c}=0.3T$, $\varphi_{0}=0.1$, $J=50g$, and $\Delta_{0}=100g$.
         (b) The average excitation fluctuations of the two membranes by applying the classical driving fields displayed in Fig. \ref{fig2} (a).
         We have chosen initial average excitation fluctuation of the first membrane as $\langle\delta b_{1}^{\dag}\delta b_{1}\rangle=1$.
         }
 \label{fig2}
\end{figure}

According to Lewis-Riesenfeld theory \cite{Jmp101458}, the solution of the differential equation $i|\dot{\Psi}(t)\rangle=M(t)|\Psi(t)\rangle$,
is $|\Psi(t)\rangle=\sum_{n}C_{n}e^{i\eta_{n}}|\phi_{n}(t)\rangle$.
We choose the eigenvector $|\phi_{0}\rangle$ with zero-eigenvalue as the evolution path, which means $C_{0}\neq 0$ and $C_{n\neq 0}=0$.
For the invariant given in Eq. (\ref{eq1-11}),
\begin{align}\label{eq1-13}
  |\phi_{0}(t)\rangle=&\cos{\varphi}\cos{\theta}|\psi_{1}\rangle-\cos{\varphi}\cos{\theta}|\psi_{2}\rangle\cr
                      &-i\sin{\varphi}|\psi_{3}\rangle.
\end{align}
Thus, the solution of the differential equation $i|\dot{\Psi}(t)\rangle=M(t)|\Psi(t)\rangle$ is obtained:
\begin{align}\label{eq1-14}
  |\Psi(t)\rangle=&C_{0}[(\cos{\varphi}\cos{\theta})\delta b_{1}-(\cos{\varphi}\sin{\theta})\delta b_{2}\cr
                  &-\frac{i}{\sqrt{2}}\sin{\varphi}(\delta a_{L}-\delta a_{R})].
\end{align}
By choosing the initial average excitation fluctuation of the first membrane as $\langle\delta b_{1}^{\dag}\delta b_{1}\rangle=1$, we have $C_{0}=1$.
Meanwhile, to transfer the fluctuation excitation from mode $b_{1}$ to mode
$b_{2}$ and to simulate classical driving fields with a finite
duration, the parameters $\theta$ and $\varphi$ should satisfy the boundaries
\begin{align}\label{eq1-15}
  \theta(t_{i})=&0, \ \theta(t_{f})=\pi/2, \cr
  \dot{\theta}(t_{i})=&0, \ \dot{\theta}(t_{f})=0, \cr
  \varphi(t_{i})=&0,\ \varphi(t_{f})=0, \cr
  \dot\varphi(t_{i})=&0,\ \dot\varphi(t_{f})=0.
\end{align}
We choose a Vitanov shape \cite{Pra80013417} for $\theta$ and a symmetric Gaussian shape for $\varphi$,
\begin{eqnarray}\label{eq1-16}
  \theta=\frac{\pi}{2[1+\exp{(-t/\tau)}]},\
  \varphi=\pi\varphi_{0}[\exp{(-t^{2}/\tau_{c}^{2})}],
\end{eqnarray}
where the time scale $\tau$ controls the effective duration of the
classical driving fields, $\tau_{c}$ is a related coefficient, and $0<\varphi_{0}<\pi/2$ is the maximum value of $\varphi$.
When the Zeno requirement is satisfied, the occupancy of the unstable ``intermediate'' state $|\phi_{3}\rangle$ is
controllable according to Eqs. (\ref{eq1-13}) and (\ref{eq1-16}). We will show in the following that there is a trade-off between
the occupancy of state $|\phi_{3}\rangle$ and the total interaction time $T$ to obtain a high-fidelity fluctuation transfer.
With suitable parameters, we display the classical driving fields versus time in Fig. \ref{fig2} (a).
Applying the classical driving fields in Fig. \ref{fig2} (a), complete fluctuation
transfer can be achieved as shown in Fig. \ref{fig2} (b) which displays the time evolution of the average excitation fluctuations
of the two membranes by solving numerically the Langevin
equations (in the absence of the decay terms) in Eq. (\ref{eq1-6}).
 \begin{figure}
 \scalebox{0.3}{\includegraphics {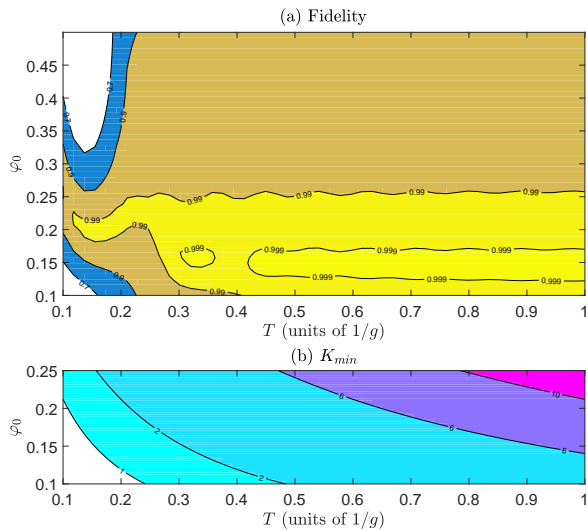}}
 \caption{
         (a) The fidelity (in absence of noises and decays) of fluctuation transfer versus evolution time $T$ and $\varphi_{0}$.
         (b) The minimum value of $K=\sqrt{2}J/\Omega$ versus $T$ and $\varphi_{0}$.
         Parameters are $\tau=0.1T$, $\tau_{c}=0.3T$, $J=50g$, and $\Delta_{0}=100g$.
         }
 \label{fig3}
\end{figure}

For the sake of convenience, we define the final (when $t=t_{f}$) average excitation fluctuation of the second
membrane $\langle\delta b_{2}^{\dag}\delta b_{2}\rangle=|\langle\Psi(t_{f})|\psi_{2}\rangle|^{2}=F$ as the fidelity of the transfer process.
Then, we plot Fig. \ref{fig3} (a) to show the fidelity $F$ versus parameters $T$ and $\varphi_{0}$.
As shown in the figure, to achieve high-fidelity fluctuation transfer with $F\geq0.999$, the shortest time required for the scheme is only $T\approx0.32/g$ when $\varphi_{0}\approx0.15$.
For $\varphi_{0}\approx0.15$, according to Eqs. (\ref{eq1-13}-\ref{eq1-16}), the maximal occupancy of the ``intermediate'' state $|\psi_{3}\rangle$ is only $|\sin{(0.15\pi)}|^2\approx0.2$.
Moreover, we find from the figure that the fidelity decreases with the increase of $\varphi_{0}$. This is because we have chosen a relatively large $\tau_{c}$ to plot the figure
so that the boundaries given in Eq. (\ref{eq1-15}) are hard to satisfy when $\varphi_{0}$ is large. For example, when $\varphi_{0}=0.4$,
we have $\varphi(t_{i})=\varphi(t_{f})\approx0.03\pi$ leading to $C_{0}\approx0.9956$ and $F\leq |C_{0}\cos{[\varphi(t_{f})]}|^4\approx0.98$.
To check if the strong coupling limit $K\rightarrow\infty$ is satisfied with the current parameters,
we plot $K_{min}$ (the minimum value of $K$) versus $T$ and $\varphi_{0}$ in Fig. \ref{fig3} (b).
It is interesting to find from Fig. \ref{fig3} that, a high fidelity $F\simeq0.99$ is achievable
even when the strong coupling limit $K\rightarrow\infty$ is not satisfied: $K_{min}\simeq2$ when $F\simeq 0.99$.
In fact, this phenomenon has been discussed in detail in Ref. \cite{Pra89033856}.
It is not so necessary to ideally satisfy the condition $K\rightarrow\infty$
in applying invariant-based inverse engineering and quantum Zeno dynamics for a time-dependent system.
We know that when the Zeno requirement is not well satisfied,
the ``intermediate'' states in the Zeno subspaces with $\epsilon\neq 0$ are no longer negligible.
However, according to Ref. \cite{Pra89033856}, under certain conditions, the intermediate states can help
to speed up the population transfer. $K\geq 2$ is enough for the current system to obtain high-fidelity fluctuation transfer.

\section{Robustness against noises and decays}
In real experiment, there is usually a stochastic kind of noise on the parameters that should be considered.
We assume that the interaction matrix $M(t)$ is perturbed by
a stochastic part $\mu M_{s}(t)$ describing amplitude noise.
The Langevin equations in form of $i|\dot{\Psi}(t)\rangle=M(t)|\Psi(t)\rangle$ thus become
$i|\dot{\Psi}(t)\rangle=[M(t)+\mu M_{s}(t)\xi(t)]|\Psi\rangle$,
where $\xi(t)=\partial_{t}W_{t}$ is heuristically the time derivative of the Brownian motion $W_{t}$.
$\xi(t)$ satisfies $\langle\xi(t)\rangle=0$ and $\langle\xi(t)\xi(t')\rangle=\delta(t-t')$ because the noise should have zero mean and the noise
at different times should be uncorrelated. Then, we define $\rho_{\xi}(t)=|\Psi_{\xi}(t)\rangle\langle\Psi_{\xi}(t)|$, and
the dynamical equation for $\rho_{\xi}(t)$ is thus given as \cite{Njp14093040}
\begin{align}\label{eq2-1}
  \dot{\rho}_{\xi}(t)=-i[M(t),\rho_{\xi}(t)]-{i\mu}[M_{s}(t),\xi\rho_{\xi}(t)].
\end{align}
After averaging over the noise, Eq. (\ref{eq2-1}) becomes
$\dot{\rho}(t)\simeq-i[M(t),\rho(t)]-{i\mu}[M_{s}(t),\langle\xi(t)\rho_{\xi}(t)\rangle]$,
where $\rho(t)=\langle\rho_{\xi}(t)\rangle$. According to Novikov's theorem in the case of white noise,
we have $\langle\xi(t)\rho_{\xi}(t)\rangle=\frac{1}{2}\langle\frac{\delta\rho_{\xi}(t')}{\delta\xi(t')}\rangle|_{t'=t}=-\frac{i\mu}{2}[M_{s}(t),\rho(t)]$,
leading to
\begin{align}\label{eq2-2}
  \dot{\rho}(t)\simeq-i[M(t),\rho(t)]-\frac{\mu^2}{2}[M_{s}(t),[M_{s}(t),\rho(t)]].
\end{align}
The fidelity of the fluctuation transfer process is thus defined as $F=\text{Tr}[\sqrt{\rho_{2}}\rho(t_f)\sqrt{\rho_{2}}]$, where $\rho_{2}=|\psi_{2}\rangle\langle\psi_{2}|$.

\begin{figure}
 \scalebox{0.31}{\includegraphics {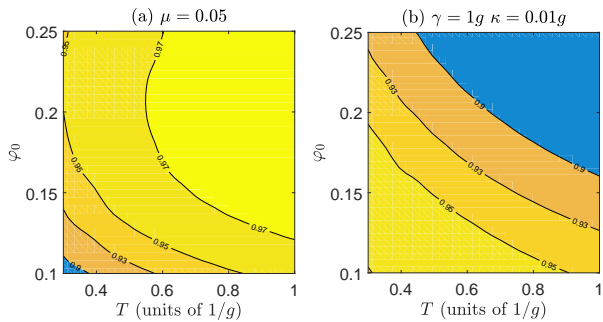}}
 \caption{
         (a) The fidelity (in absence of decays) of the scheme versus $\varphi_{0}$ and $T$ when the amplitude noises are considered.
         (b) The fidelity (in absence of noises) of the scheme versus $\varphi_{0}$ and $T$  when the cavity decays are considered.
             $0.1\leq\varphi_{0}\leq 0.25$ and $0.3/g\leq T\leq 1/g$ are chosen according to Fig. \ref{fig3} for a high-fidelity transfer.
             Parameters are $\tau=0.1T$, $\tau_{c}=0.3T$, $J=50g$, and $\Delta_{0}=100g$.
         }
 \label{fig4}
\end{figure}

In the current scheme, we consider independent amplitude noise in Rabi frequency $\Omega_{L}(t)$ as well as in $\Omega_{R}(t)$ with the same intensity $\mu$.
The amplitude noise in other parameters affect the scheme very weakly thus can be ignored.
In this case, the master equation is
\begin{align}\label{eq2-3}
  \dot{\rho}(t)=&-{i}[M(t),\rho(t)]-{\mu^2}[M_{aL}(t),[M_{aL}(t),\rho(t)]]\cr
                                                    &-{\mu^2}[M_{aR}(t),[M_{aR}(t),\rho(t)]],
\end{align}
where
\begin{align}\label{eq2-4}
  M_{aL}&=\frac{g_{1}\Omega_{L}}{-\Delta_{0}}
        \left(
              \begin{array}{ccccc}
                0 & 0 & 0 & 1 & 0 \\
                0 & 0 & 0 & 0 & 0 \\
                0 & 0 & 0 & 0 & 0 \\
                1 & 0 & 0 & 0 & 0 \\
                0 & 0 & 0 & 0 & 0
              \end{array}
        \right),\cr
  M_{aR}&=\frac{g_{2}\Omega_{R}}{-\Delta_{0}}
         \left(
              \begin{array}{ccccc}
                0 & 0 & 0 & 0 & 0 \\
                0 & 0 & 0 & 0 & 0 \\
                0 & 0 & 0 & 0 & 1 \\
                0 & 0 & 0 & 0 & 0 \\
                0 & 0 & 1 & 0 & 0
              \end{array}
        \right).
\end{align}
Assuming the intensity of noise is $\mu=0.05$, the sensitivity with respect to amplitude-noise error of the population transfer is shown in Fig. \ref{fig4} (a).
The sensitivity with respect to amplitude-noise error decreases with the increases of the $T$ and $\varphi_{0}$.
As we know, the adiabatic condition for the current system is satisfied better with a longer interaction time,
and an adiabatic process is generally much more robust against noises than a non-adiabatic process.
That is why we find from the figure that shortening the interaction time increases the sensitivity with respect to amplitude-noise error.

In the following, we would like to analyze the influence of the decays on the scheme.
The non-Hermitian terms in Eq. (7) should be taken into account in the numerical simulation in this case.
Without loss of generality, we set $\gamma_{j}=j$ and $\gamma_{m,k}=\kappa$.
Then, with Eq. (\ref{eq1-7}), we plot Fig. \ref{fig4} (b) to show the fidelity $F$ versus $J$ and $T$ when decays are considered.
The influence of decays increase with the increase of $T$ and $\varphi_{0}$ as shown in the figure.
This result is contradictory to that of Fig. \ref{fig4} (a) as large $T$ and $\varphi_{0}$ are required for the robustness against noises.
In practice, if the experimental facility allows for weak noises in the driving fields, one can achieve the energy fluctuation transfer between
membranes with a high fidelity $\geq0.95$ by choosing $T\leq 0.5/g$ and $\varphi_{0}\leq 0.15$.

\section{Conclusion}
We have presented a scheme in an optomechanical cavity system to
transfer the average fluctuation of excitation from one membrane to
the other. The cavity is divided into three subcavities by placing
two membranes, and the subcavities couple to each other
via tunneling through the membranes. Each of the subcavities is
considered to decay independently. By invariant-based inverse
engineering, we have designed time-dependent classical driving fields in
counterintuitive sequences to guide the system to evolve along a
non-adiabatic path to realize the transfer. The evolution path is
parametrized so that one can control the occupancies of the
unstable intermediate states to suppress the influence of
decays. We have numerically shown the choices of the optimal
parameters to achieve a high-fidelity fluctuation transfer.
Moreover, by numerical simulation, we have further
investigated the influence of noises and decays; the result shows
the present scheme is robust against the amplitude noises and decays
of the membranes and the cavity modes. Thus, we hope the present
scheme might benefit quantum information science based on
quantum optomechanics.

\section*{ACKNOWLEDGMENT}

The authors thank Dr. Chang-Sheng Hu for valuable discussions. This work was supported by the National Natural Science Foundation
of China under Grants No. 11575045, No. 11374054, No. 11675046, and No. 11747011.

\begin{appendix}
\section*{APPENDIX A: Derivation of system Hamiltonian in Eq. (2)}
The dynamical Hamiltonian for the system in Fig. \ref{fig1} without applying the classical driving fields is ($\hbar=1$)
\begin{align}\label{eqR1}
  H=&H_{0}+H_{J},\cr
  H_{0}=&\sum_{j=L,M,R}\omega_{c,j}a_{j}^{\dag}a_{j},\cr
  H_{J}=&-(J_{1}a_{L}^{\dag}a_{M}+J_{2}a_{M}^{\dag}a_{R}+H.c.), \cr
  \tag{A1}
\end{align}
Then, we send three monochromatic waves $\Omega_{L}$, $\Omega_{R}$, and $\Omega_{M}$
along the cavity axis to drive the modes $a_{L}$, $a_{R}$, and $a_{M}$, respectively. The driving Hamiltonian reads
\begin{align}\label{eqR21}
  H_{D}=&\Omega_{L}a_{L}e^{i\omega_{l}t}+\Omega_{R}a_{R}e^{i\omega_{l}t} \cr
        &+\Omega_{M}a_{M}e^{i\omega_{l}'t}+H.c.,
  \tag{A2}
\end{align}
Caused by the optical radiation pressure, the membranes 1 and 2 will move with displacements $q_{1}$ and $q_{2}$, respectively (as shown in Fig. \ref{fig1}).
Thus, the resonance frequencies of the cavity modes will be changed \cite{Prl103100402,Njp16033023,Pra92013822} and $H_{0}$ in Eq. (\ref{eqR1}) should be modified as
\begin{align}\label{eqR3}
  H_{0}'=&\frac{\omega_{c,L}}{1+\frac{q_{1}}{\zeta_{L}\mathcal{L}}}a_{L}^{\dag}a_{L}+\frac{\omega_{c,R}}{1-\frac{q_{2}}{\zeta_{R}\mathcal{L}}}a_{R}^{\dag}a_{R} \cr
         &+\frac{\omega_{c,M}}{1+\frac{q_{2}-q_{1}}{\zeta_{M}\mathcal{L}}}a_{M}^{\dag}a_{M},
\tag{A3}
\end{align}
where
\begin{align}\label{eqR4}
  \zeta_{L}=&\frac{\omega_{c,L}}{\omega_{c,L}+\omega_{c,R}+\omega_{c,M}}\cr
  \zeta_{M}=&\frac{\omega_{c,M}}{\omega_{c,L}+\omega_{c,R}+\omega_{c,M}},\cr
  \zeta_{R}=&\frac{\omega_{c,R}}{\omega_{c,L}+\omega_{c,R}+\omega_{c,M}}.
\tag{A4}
\end{align}
Then, considering that $q_{1},q_{2}\ll \zeta_{j}{\mathcal{L}}$ ($j=L,M,R$), we have
\begin{align}\label{eqR5}
  \frac{\omega_{c,L}}{1+\frac{q_{1}}{\zeta_{L}\mathcal{L}}}\approx&(1-\frac{q_{1}}{\zeta_{L}\mathcal{L}})\omega_{c,L}\cr
                                                           =&\omega_{c,L}-\frac{q_{1}}{\mathcal{L}}\Xi,\cr
  \frac{\omega_{c,R}}{1-\frac{q_{2}}{\zeta_{R}\mathcal{L}}}\approx&(1+\frac{q_{2}}{\zeta_{R}\mathcal{L}})\omega_{c,R}\cr
                                                           =&\omega_{c,R}+\frac{q_{2}}{\mathcal{L}}\Xi,\cr
  \frac{\omega_{c,M}}{1+\frac{q_{2}-q_{1}}{\zeta_{M}\mathcal{L}}}\approx&(1-\frac{q_{2}-q_{1}}{\zeta_{M}\mathcal{L}})\omega_{c,M}\cr
                                                           =&\omega_{c,M}-\frac{q_{2}-q_{1}}{\mathcal{L}}\Xi,
\tag{A5}
\end{align}
Substituting Eq. (\ref{eqR5}) into Eq. (\ref{eqR3}), we obtain
\begin{align}\label{eqR6}
  H_{0}'\approx &\sum_{j=L,M,R}\omega_{c,j}a_{j}^{\dag}a_{j}-(\frac{\Xi}{\mathcal{L}}a_{L}^{\dag}a_{L}-\frac{\Xi}{\mathcal{L}}a_{M}^{\dag}a_{M})q_{1} \cr
                 &+(\frac{\Xi}{\mathcal{L}}a_{R}^{\dag}a_{R}-\frac{\Xi}{\mathcal{L}}a_{M}^{\dag}a_{M})q_{2}.
\tag{A6}
\end{align}
Meanwhile, the free Hamiltonian for the oscillators is
\begin{align}\label{eqR7}
  H_{00}=\sum_{k=1,2}\frac{p_{k}^{2}}{2\mu_{k}}+\frac{1}{2}\mu_{k}{\omega_{m,k}^{2}q_{k}^2},
\tag{A7}
\end{align}
where $p_{k}$ is the $k$th momentum operator of the movable membranes and $\mu_{k}$ is the corresponding membrane mass.
It will be more convenient to write the position and momentum operators in terms of the annihilation operator ($b_{k}$) and
creation operator ($b_{k}^{\dag}$) as $q_{k}=\sqrt{\frac{1}{2\mu_{k}\omega_{m,k}}}(b_{k}+b_{k}^{\dag})$ and $p_{k}=i\sqrt{\frac{\mu_{k}\omega_{m,k}}{2}}(b_{k}^{\dag}-b_{k})$, respectively.
$b_{k}$ and $b_{k}^{\dag}$ satisfy $[b_{k},b_{k}^{\dag}]=1$. Then, Eq. (\ref{eqR7}) becomes
\begin{align}\label{eqR8}
  H_{00}=\sum_{k=1,2}\omega_{m,k}b_{k}^{\dag}b_{k}.
\tag{A8}
\end{align}
Note that the transmission coefficient $J_{1}$ ($J_{2}$) between the left (right) and middle cavity modes through first
(second) membrane mainly depends on the property of the membranes so that the Hamiltonian $H_{J}$ remains unchanged.
In this case, by choosing $g_{1}=\frac{\Xi}{\mathcal{L}}\sqrt{\frac{1}{2\mu_{1}\omega_{m,1}}}$, $g_{2}=\frac{\Xi}{\mathcal{L}}\sqrt{\frac{1}{2\mu_{2}\omega_{m,2}}}$, and
\begin{align}\label{eqR9}
  H_{I}=&-g_{1}(a_{L}^{\dag}a_{L}-a_{M}^{\dag}a_{M})(b_{1}+b_{1}^{\dag}) \cr
                 &+g_{2}(a_{R}^{\dag}a_{R}-a_{M}^{\dag}a_{M})(b_{2}+b_{2}^{\dag}), \cr
  H_{0}=&H_{0}'+H_{00}-H_{I}          \cr
       =&\sum_{j=L,M,R}\omega_{c,j}a_{j}^{\dag}a_{j}+\sum_{k=1,2}\omega_{m,k}b_{k}^{\dag}b_{k},
\tag{A9}
\end{align}
the Hamiltonian in Eq. (2) is obtained.

\begin{widetext}
\section*{APPENDIX B: Derivation of effective Hamiltonian in Eq. (4)}
According to Refs. \cite{Quantumoptics,Pra96023837}, by using the input-output formalism, we can obtain Langevin equations for the relevant
operators in Eq. (\ref{eq1-2}) as
\begin{align} \label{eqa-1}
  \dot{a}_{L}=&-(\gamma_{L}/2+i\Delta_{L})a_{L}+iJ_{1}a_{M}+ig_{1}a_{L}(b_{1}+b_{1}^{\dag})-i\Omega_{L}-\sqrt{\gamma_{L}}a_{L}^{\text{in}}, \cr
  \dot{a}_{R}=&-(\gamma_{R}/2+i\Delta_{R})a_{R}+iJ_{2}a_{M}-ig_{2}a_{R}(b_{2}+b_{2}^{\dag})-i\Omega_{R}-\sqrt{\gamma_{R}}a_{R}^{\text{in}}, \cr
  \dot{a}_{M}=&-(\gamma_{M}/2+i\Delta_{M})a_{M}+iJ_{1}a_{L}+iJ_{2}a_{R}-ig_{1}a_{M}(b_{1}+b_{1}^{\dag})+ig_{2}a_{M}(b_{2}+b_{2}^{\dag})-i\Omega_{M}-\sqrt{\gamma_{M}}a_{M}^{\text{in}}, \cr
  \dot{b}_{1}=&-(\gamma_{m,1}/2+i\omega_{m,1})b_{1}+ig_{1}(a_{L}^{\dag}a_{L}-a_{M}^{\dag}a_{M})+\sqrt{\gamma_{m}}b_{1}^{\text{in}}, \cr
  \dot{b}_{2}=&-(\gamma_{m,2}/2+i\omega_{m,2})b_{2}-ig_{2}(a_{R}^{\dag}a_{R}-a_{M}^{\dag}a_{M})+\sqrt{\gamma_{m}}b_{2}^{\text{in}},
  \tag{B1}
\end{align}
where $\gamma_{j}$ is the decay rate of the $j$th mode of the cavity and $\gamma_{m,k}$ ($k=1,2$)
is the dissipation rate of the $k$th membrane, $a_{j}^{\text{in}}$ and $b_{k}^{\text{in}}$
are noise operators \cite{Quantumoptics} satisfying
\begin{align}\label{eqa-2}
  \langle a_{j}^{\text{in}}(t)a_{j}^{\dag \text{in}}(t')\rangle=&\delta(t-t'), \cr
  \langle a_{j}^{\dag\text{in}}(t)a_{j}^{\text{in}}(t')\rangle=&0, \cr
  \langle b_{k}^{\text{in}}(t)b_{k}^{\text{in}\dag}(t')\rangle=&(\bar{n}_{\text{th}}+1)\delta(t-t'), \cr
  \langle b_{k}^{\text{in}\dag}(t)b_{k}^{\text{in}}(t')\rangle=&(\bar{n}_{\text{th}})\delta(t-t'), \cr
  \tag{B2}
\end{align}
where $\bar{n}_{\text{th}}=\{\exp[\hbar\omega_{m,k}/(k_{B}T)]\}^{-1}$ is the mean thermal
excitation number in the bath, interacting with the mechanical
oscillator with frequency $\omega_{m,k}$ at an equilibrium temperature $T$
and $k_{B}$ is the Boltzmann constant \cite{Pra96023837}.
Then, by using the standard linearization procedure \cite{Rmp861391} to expand the bosonic operators as a sum of
the average values and the zero-mean fluctuation as $a_{j}\rightarrow\alpha_{j}+\delta a_{j}$ and $b_{k}\rightarrow\beta_{k}+\delta b_{k}$,
we obtain the following equations for the average of the operators
\begin{align}\label{eqa-3}
  \dot{\alpha}_{L}=&-(\gamma_{L}/2+i\Delta_{L}')\alpha_{L}+iJ_{1}\alpha_{M}-i\Omega_{L}, \cr
  \dot{\alpha}_{R}=&-(\gamma_{R}/2+i\Delta_{R}')\alpha_{R}+iJ_{2}\alpha_{M}-i\Omega_{R}, \cr
  \dot{\alpha}_{M}=&-(\gamma_{M}/2+i\Delta_{M}')\alpha_{M}+iJ_{1}\alpha_{L}+iJ_{2}\alpha_{R}-i\Omega_{M}, \cr
  \dot{\beta}_{1}=&-(\gamma_{m,1}/2+i\omega_{m,1})\beta_{1}+ig_{1}(|\alpha_{L}|^{2}-|\alpha_{M}|^{2}), \cr
  \dot{\beta}_{2}=&-(\gamma_{m,2}/2+i\omega_{m,2})\beta_{2}-ig_{2}(|\alpha_{R}|^{2}-|\alpha_{M}|^{2}),
  \tag{B3}
\end{align}
with
\begin{align}\label{eqa-4}
  \Delta_{L}'=&\Delta_{L}-g_{1}\text{Re}(\beta_{1}), \cr
  \Delta_{R}'=&\Delta_{R}+g_{2}\text{Re}(\beta_{2}), \cr
  \Delta_{M}'=&\Delta_{M}+g_{1}\text{Re}(\beta_{1})-g_{2}\text{Re}(\beta_{2}).
  \tag{B4}
\end{align}
Solving the equations $\dot{\alpha}_{j}=0$, the steady-state solutions for $\alpha_{j}$ are obtained as
\begin{align}\label{eqa-5}
  \alpha_{L}=&\frac{\Omega_{L}-J_{1}\alpha_{M}}{-\Delta_{L}'+i\gamma_{L}/2}, \cr
  \alpha_{R}=&\frac{\Omega_{R}-J_{2}\alpha_{M}}{-\Delta_{R}'+i\gamma_{R}/2}, \cr
  \alpha_{M}=&\frac{\Omega_{M}-J_{1}\alpha_{L}-J_{2}\alpha_{R}}{-\Delta_{M}'+i\gamma_{M}/2}.
  \tag{B5}
\end{align}
Meanwhile, for the part of fluctuations, the Langevin equations are
\begin{align}\label{eqa-6}
  \delta\dot{a}_{L}=&-(\gamma_{L}/2+i\Delta_{L}')\delta\alpha_{L}+iJ_{1}\delta\alpha_{M}+ig_{1}\alpha_{L}(\delta b_{1}+\delta b_{1}^{\dag})+\sqrt{\gamma_{L}}\delta a_{L}^{\text{in}}, \cr
  \delta\dot{a}_{R}=&-(\gamma_{R}/2+i\Delta_{R}')\delta\alpha_{R}+iJ_{2}\delta\alpha_{M}+ig_{2}\alpha_{R}(\delta b_{2}+\delta b_{2}^{\dag})+\sqrt{\gamma_{R}}\delta a_{R}^{\text{in}}, \cr
  \delta\dot{a}_{M}=&-(\gamma_{M}/2+i\Delta_{M}')\delta\alpha_{M}+iJ_{1}\delta a_{L}+iJ_{2}\delta\alpha_{R}
                     -ig_{1}\alpha_{M}(\delta b_{1}+\delta b_{1}^{\dag})+ig_{2}\alpha_{M}(\delta b_{2}+\delta b_{2}^{\dag})+\sqrt{\gamma_{M}}\delta a_{M}^{\text{in}}, \cr
  \delta\dot{b}_{1}=&-(\gamma_{m,1}/2+i\omega_{m,1})\delta\beta_{1}+ig_{1}[\alpha_{L}(\delta a_{L}+\delta a_{L}^{\dag})-\alpha_{M}(\delta a_{M}+\delta a_{M}^{\dag})]+\sqrt{\gamma_{m,1}}\delta b_{1}^{\text{in}}, \cr
  \delta\dot{b}_{2}=&-(\gamma_{m,2}/2+i\omega_{m,2})\delta\beta_{2}-ig_{2}[\alpha_{R}(\delta a_{R}+\delta a_{R}^{\dag})-\alpha_{M}(\delta a_{M}+\delta a_{M}^{\dag})]+\sqrt{\gamma_{m,2}}\delta b_{2}^{\text{in}}.
  \tag{B6}
\end{align}
Equations (\ref{eqa-6}) can be derived from the following
linearized Hamiltonian,
\begin{align}\label{eqa-7}
  H=&\sum_{j=L,M,R}\Delta_{j}'\delta a_{j}^{\dag}a\delta a_{j}+\sum_{k=1,2}\omega_{m,k}\delta b_{k}^{\dag}b_{k} \cr
    &-g_{1}(\alpha_{L}^{*}\delta a_{L}+\alpha_{L}\delta a_{L}^{\dag}-\alpha_{M}^{*}\delta a_{M}-\alpha_{M}\delta a_{M}^{\dag})(\delta b_{1}+\delta b_{1}^{\dag})\cr
    &+g_{2}(\alpha_{R}^{*}\delta a_{R}+\alpha_{R}\delta a_{R}^{\dag}-\alpha_{M}^{*}\delta a_{M}-\alpha_{M}\delta a_{M}^{\dag})(\delta b_{2}+\delta b_{2}^{\dag})\cr
    &-(J_{1}\delta a_{L}^{\dag}\delta a_{M}+J_{2}\delta a_{R}^{\dag}\delta a_{M}+H.c.).
    \tag{B7}
\end{align}
Then, choosing $H_{0}=\sum_{j=L,M,R}\Delta_{j}'\delta a_{j}^{\dag}a\delta a_{j}+\sum_{k=1,2}\omega_{m,k}\delta b_{k}^{\dag}b_{k}$ as the unperturbed Hamiltonian,
performing the unitary transformation $U=\exp{(-iH_{0}t)}$, we obtain the Hamiltonian in the interaction picture as
\begin{align}\label{eqa-8}
  H_{i}=U^{\dag}HU=&-g_{1}(\alpha_{L}^{*}\delta a_{L}e^{-i\Delta_{L}'t}+\alpha \delta a_{L}^{\dag}e^{i\Delta_{L}'t})(\delta b_{1}e^{-i\omega_{m,1}t}+\delta b_{1}^{\dag}e^{i\omega_{m,1}t})\cr
                   &+g_{1}(\alpha_{M}^{*}\delta a_{M}e^{-i\Delta_{M}'t}+\alpha \delta a_{M}^{\dag}e^{i\Delta_{M}'t})(\delta b_{1}e^{-i\omega_{m,1}t}+\delta b_{1}^{\dag}e^{i\omega_{m,1}t})\cr
                   &+g_{2}(\alpha_{R}^{*}\delta a_{R}e^{-i\Delta_{R}'t}+\alpha \delta a_{R}^{\dag}e^{i\Delta_{R}'t})(\delta b_{2}e^{-i\omega_{m,2}t}+\delta b_{2}^{\dag}e^{i\omega_{m,2}t})\cr
                   &-g_{2}(\alpha_{M}^{*}\delta a_{M}e^{-i\Delta_{M}'t}+\alpha e^\delta a_{M}^{\dag}{i\Delta_{M}'t})(\delta b_{2}e^{-i\omega_{m,2}t}+\delta b_{2}^{\dag}e^{i\omega_{m,2}t})\cr
                   &-[J_{1}\delta a_{L}^{\dag}\delta a_{M}e^{i(\Delta_{L}'-\Delta_{M}')t}+J_{2}\delta a_{R}^{\dag}a_{R}e^{i(\Delta_{R}'-\Delta_{M}')t}].
                   \tag{B8}
\end{align}
Considering the weak-coupling condition that $\Delta_{j}',\omega_{m,k}\gg |g_{1}\alpha_{M}|,|g_{2}\alpha_{R}|,|g_{2}\alpha_{M}|$, we can perform the
rotating-wave approximation, and we obtain the effective Hamiltonian
\begin{align}\label{eqa-9}
  H_{i}\simeq&-g_{1}[\alpha_{L}^{*}\delta a_{L}\delta b_{1}^{\dag}e^{i(\omega_{m,1}-\Delta_{L}')t}]
              +g_{1}[\alpha_{M}^{*}\delta a_{M}\delta b_{1}^{\dag}e^{i(\omega_{m,1}-\Delta_{M}')t}] \cr
             &+g_{2}[\alpha_{R}^{*}\delta a_{R}\delta b_{2}^{\dag}e^{i(\omega_{m,2}-\Delta_{R}')t}]
              -g_{2}[\alpha_{M}^{*}\delta a_{M}\delta b_{2}^{\dag}e^{i(\omega_{m,2}-\Delta_{M}')t}] \cr
             &-[J_{1}\delta a_{L}^{\dag}\delta a_{M}e^{i(\Delta_{L}'-\Delta_{M}')t}+J_{2}\delta a_{R}^{\dag}a_{R}e^{i(\Delta_{R}'-\Delta_{M}')t}]+H.c..
             \tag{B9}
\end{align}
Obviously, when we choose $\Delta_{j}'=\omega_{m,k}=\Delta_{0}$, the final effective Hamiltonian in Eq. (\ref{eq1-3}) is obtained.
\end{widetext}

\end{appendix}

\end{document}